\documentclass{mymodwebofc}
\pdfoutput=1

\usepackage[varg]{txfonts}   
\usepackage{booktabs}
\usepackage{mathpazo} 
\usepackage[scaled]{helvet} 
\usepackage{courier} 
\normalfont
\usepackage[T1]{fontenc}

\woctitle{ }

\newcommand{\TeV}{\ensuremath{{\text{TeV}}}\xspace}
\newcommand{\GeV}{\ensuremath{{\text{GeV}}}\xspace}
\newcommand{\MeV}{\ensuremath{{\text{MeV}}}\xspace}

\newcommand{\micron}{\ensuremath{\mu\text{m}}\xspace}
\newcommand{\rad}{\ensuremath{{\text{rad}}}\xspace}
\newcommand{\fb}{\ensuremath{{\text{fb}^{-1}}}\xspace}
\newcommand{\ps}{\ensuremath{{\text{ps}^{-1}}}\xspace}
\newcommand{\pt}{\ensuremath{p_{\text{T}}}\xspace}

\newcommand{\pT}{\ensuremath{p_{\text{T}}}\xspace}

\newcommand{\chisq}{\ensuremath{\chi^{2}}\xspace}

\newcommand{\eff}{\ensuremath{\varepsilon_{tag}}\xspace}
\newcommand{\wtag}{\ensuremath{\omega}\xspace}

\newcommand{\ptag}{\ensuremath{\mathcal{P}_{tag}}\xspace}

\newcommand{\kaon}{\ensuremath{\text{K}}\xspace}
\newcommand{\kaonp}{\ensuremath{\text{K}^{+}}\xspace}
\newcommand{\kaonm}{\ensuremath{\text{K}^{-}}\xspace}

\newcommand{\muon}{\ensuremath{\mu}\xspace}
\newcommand{\muonp}{\ensuremath{\mu^{+}}\xspace}
\newcommand{\muonm}{\ensuremath{\mu^{-}}\xspace}
\newcommand{\ele}{\ensuremath{e}\xspace}

\newcommand{\bquark}{\ensuremath{\text{b}}\xspace}

\newcommand{\cquark}{\ensuremath{\text{c}}\xspace}
\newcommand{\cquarkbar}{\ensuremath{\bar{\text{c}}}\xspace}
\newcommand{\squark}{\ensuremath{\text{s}}\xspace}

\newcommand{\B}{\ensuremath{\text{B}}\xspace}
\newcommand{\BsL}{\ensuremath{\text{B}_{L}}\xspace}
\newcommand{\BsH}{\ensuremath{\text{B}_{H}}\xspace}

\newcommand{\Bs}{\ensuremath{\text{B}^{0}_{\text{s}}}\xspace}

\newcommand{\Bz}{\ensuremath{\text{B}^{0}}\xspace}

\newcommand{\Bp}{\ensuremath{\text{B}^{+}}\xspace}

\newcommand{\Bc}{\ensuremath{\text{B}_{\text{c}}}\xspace}

\newcommand{\Jpsi}{\ensuremath{\text{J}/\psi}\xspace}

\newcommand{\PhiTenTwenty}{\ensuremath{\phi\left(1020\right)}\xspace}

\newcommand{\Lb}{\ensuremath{\Lambda_{\text{b}}}\xspace}
\newcommand{\BpJpsiK}{\ensuremath{\Bp \! \to \Jpsi\,\text{K}^{+}}\xspace}

\newcommand{\BsJpsiPhi}{\ensuremath{\Bs \! \to \Jpsi\,\phi}\xspace}

\newcommand{\BsJpsiMMPhiKK}{\ensuremath{\Bs \! \to \Jpsi\left(\muon\muon\right)\,\phi\left(\kaon\kaon\right)}\xspace}
\newcommand{\DeltaGammaS}{\ensuremath{\Delta \! \Gamma_{\text{s}}}\xspace}
\newcommand{\DeltaMs}{\ensuremath{\Delta m_{\text{s}}}\xspace}
\newcommand{\PhiS}{\ensuremath{\phi_{\text{s}}}\xspace}
\newcommand{\ie}{i.e.\xspace}

\newcommand{\stat}{\ensuremath{\text{(stat.)}}\xspace}
\newcommand{\syst}{\ensuremath{\text{(syst.)}}\xspace}

\begin{document}

\title{Measurement of CP Violation in \BsJpsiPhi decays with the CMS detector.}
\author{ Jacopo Pazzini\inst{1} on behalf of the CMS collaboration.}
\institute{Universit\`a di Padova, INFN sezione di Padova}
\subtitle{Presented at the 8th International Workshop on the CKM Unitarity Triangle (CKM 2014), Vienna, Austria, September 8-12, 2014}

\abstract
{
  The CP-violating weak phase \PhiS and the decay width difference \DeltaGammaS of \Bs mesons are measured by the CMS experiment at the LHC using a data sample of \BsJpsiMMPhiKK decays. 
  The analysed dataset corresponds to an integrated luminosity of about $20~\fb$ collected in $pp$ collisions at a centre-of-mass energy $\sqrt{s} = 8~\TeV$. 
  A total of $49\,000$ reconstructed \Bs decays are used to extract the values \PhiS and \DeltaGammaS by performing a time-dependent and flavour-tagged angular analysis of the $\mu^+\mu^-\kaon^+\kaon^-$ final state. 
  The weak phase is measured to be $\PhiS = -0.03 \pm 0.11~\stat \pm 0.03~\syst~\rad$, and the decay width difference between the \Bs mass eigenstates is $\DeltaGammaS = 0.096 \pm 0.014~\stat \pm 0.007~\syst~\ps$.  
}

\maketitle

\section{Introduction} \label{sec:int}
Neutral \B mesons are subject to mixing, \ie, oscillations from particle to antiparticle through flavour changing neutral currents quark transition that change the meson flavour by two units, $\Delta\B = 2$.
The \Bs mixing is characterized by the mass difference \DeltaMs and by the decay width difference \DeltaGammaS between the heavy (\BsH) and light (\BsL) mass eigenstates.
A CP-violating phase \PhiS arise from the interference between direct \Bs meson decays into a $\bquark\to\cquark\cquarkbar\squark$ CP eigenstate, and decays mediated by mixing to the same final state. 
The two corresponding phases $\PhiS^D$ and $\PhiS^M$ depend on the convention of the CKM matrix parameterization. 
However, the difference \PhiS is phase-independent, and neglecting penguin diagram contributions, it is related to the elements of the CKM matrix, as: $\PhiS \simeq -2\beta_{\text{s}}$, where $\beta_{\text{s}}=\arg(-V_{ts}V^*_{tb}/V_{cs}V^*_{cb})$. 
A value of $\PhiS \simeq 2\beta_\mathrm{s} = 0.0363^{+0.0016}_{-0.0015}~\rad$, is predicted by the standard model (SM), determined via a global fit to experimental data~\cite{bib:Charles}.
Since the value is small and precisely predicted, any deviation of the measured value would be particularly interesting as a possible hint of physics beyond the SM, contributing in the \Bs mixing.
The decay width difference \DeltaGammaS is predicted to be non-zero in the SM, and the theoretical prediction, assuming no new physics in \Bs mixing, is $\DeltaGammaS = 0.087\pm 0.021~\ps$~\cite{bib:Lenz}.
In this measurement the \BsJpsiMMPhiKK decay channel has a non-definite CP final state, and an angular analysis is therefore applied to disentangle the CP-odd and CP-even components. 
A time-dependent angular analysis is performed with the CMS detector~\cite{bib:Adolphi} by measuring the decay angles of the final state particles $\muonp\muonm\kaonp\kaonm$, and the proper decay length of the \Bs. 
In this measurement the transversity basis is used~\cite{bib:Dighe_99}.
The angles $\theta_T$ and $\varphi_T$ are the polar and azimuthal angles of the $\muonp$ in the rest frame of the \Jpsi, respectively, where the $x$ axis is defined by the direction of the \PhiTenTwenty meson in the \Jpsi rest frame, and the $x$-$y$ plane is defined by the decay plane of the $\PhiTenTwenty \! \to \! \kaonp \kaonm$. 
The helicity angle $\psi_T$ is the angle of the \kaonp in the \PhiTenTwenty rest frame with respect to the negative \Jpsi momentum direction.
The differential decay rate of the \BsJpsiPhi in terms of proper decay length and angular variables is represented according to Ref.~\cite{bib:Dighe_96}, as:
\begin{equation}
\frac{d^4\Gamma(\Bs)}{d\Theta dct} = f(\Theta,\alpha,ct)\propto \sum^{10}_{i=1}O_i(\alpha,ct)\cdot g_i(\Theta),
\label{eqnarray:decayrate}
\end {equation}
where $O_i$ are the time-dependent functions, $g_i$ are the angular functions, $\Theta$ represents the angles, and $ct$ represents the proper decay length of the \Bs meson.
A detailed description of the signal model can be found in Ref.~\cite{bib:PAS}.
\section{Event selection and simulated samples}\label{sec:evsel}
\begin{figure}[htb]
  \begin{center}
    \includegraphics[width=.45\textwidth]{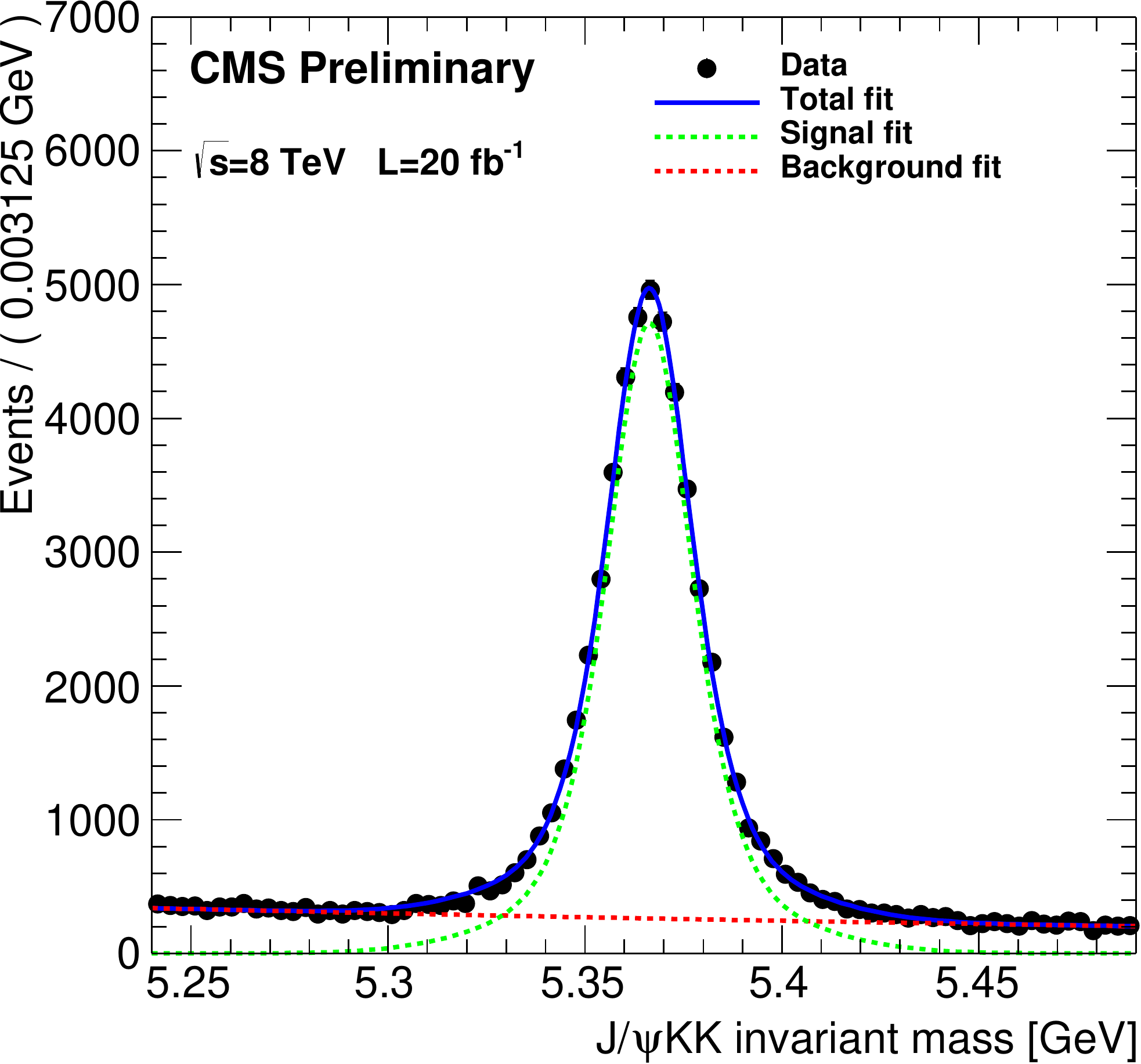}
    \caption{The mass distribution of the $\Jpsi\kaon\kaon$ candidates. The full line is a fit to the data (solid markers), the dashed green line is the fitted signal and the dashed red line is the fitted background.}
    \label{fig:mass}
  \end{center}
\end{figure}

The analized events are selected with a trigger optimized for the detection of \bquark-hadrons decaying to $\Jpsi(\muonp\muonm)$, with a dimuon invariant mass within the range $[2.9 - 3.3]~\GeV$, and transverse momentum (\pt) greater than $6.9~\GeV$.
The muon trajectories are fit to a common decay vertex, and the transverse decay length significance $L_{xy}/\sigma_{L_{xy}}$ is required to be greater than three, where $L_{xy}$ is the distance between the primary and secondary vertex in the transverse plane, and $\sigma_{L_{xy}}$ is its uncertainty. 
The vertex fit probability is required to be larger than 15\%.
Offline selection criteria requires the \Jpsi to be reconstructed using muons with a transverse momentum greater than $4~\GeV$.
The dimuon invariant mass is required to lie within $150~\MeV$ from the world-average \Jpsi mass value~\cite{bib:Beringer}.
Candidate \PhiTenTwenty mesons are reconstructed from pairs of oppositely charged tracks with $\pt > 0.7~\GeV$, after excluding the muon candidate tracks forming the \Jpsi. 
Each selected track is assumed to be a kaon and the invariant mass the candidate \PhiTenTwenty is required to be within $10~\MeV$ of the world average meson mass~\cite{bib:Beringer}.
\Bs candidates are formed by combining a \Jpsi with a \PhiTenTwenty candidate. 
The two muons and the two kaons are fitted with a combined vertex and kinematic fit, with a constraint of the dimuon invariant mass to be the nominal \Jpsi mass~\cite{bib:Beringer}. 
A \Bs candidate is retained if the $\Jpsi\,\phi$ pair has an invariant mass between 5.20 and 5.65~\GeV and the \chisq vertex fit probability is larger than 2\%. 
For each selected event the primary vertex which minimises the angle between the flight direction and the momentum of the \Bs candidate is selected.
For events with more than one \Bs candidate, the candidate with the highest vertex fit probability is selected. 
Simulated events are produced using the PYTHIA 6.4 Monte Carlo event generator and EVTGEN simulation package.
%
The generated events are then passed through a full CMS detector simulation using the GEANT package. 
Simulated \BsJpsiPhi samples, validated through comparison with the data, are used to determine the signal reconstruction efficiencies, and to estimate the background components in the signal mass window.
The angular efficiency correction $\epsilon(\Theta)$ is obtained from simulations with a three-dimensional function of the angular variables in order to take into account the correlation between the angular observables.
The proper decay length, $ct$, is required to be larger than $200~\micron$ in order to avoid a lifetime bias due to the turn-on curve of the trigger efficiency.
The main background for the \BsJpsiPhi decays originates from non-prompt \Jpsi arising from the decay of b-hadrons, such as \Bz, \Bp and \Lb.  
The \Bc cross section is expected to be very small and therefore \Bc decays are not considered. 
The \Lb contribution to the selected events is also found to be very small and the mass distribution in the selected mass range is observed to be flat. 
The mass distribution of the signal region is shown in Fig.~\ref{fig:mass}.

\section{Flavour Tagging}\label{sec:flavourtag}
\begin{figure}[htb]
\begin{center}
\includegraphics[width=0.45\textwidth]{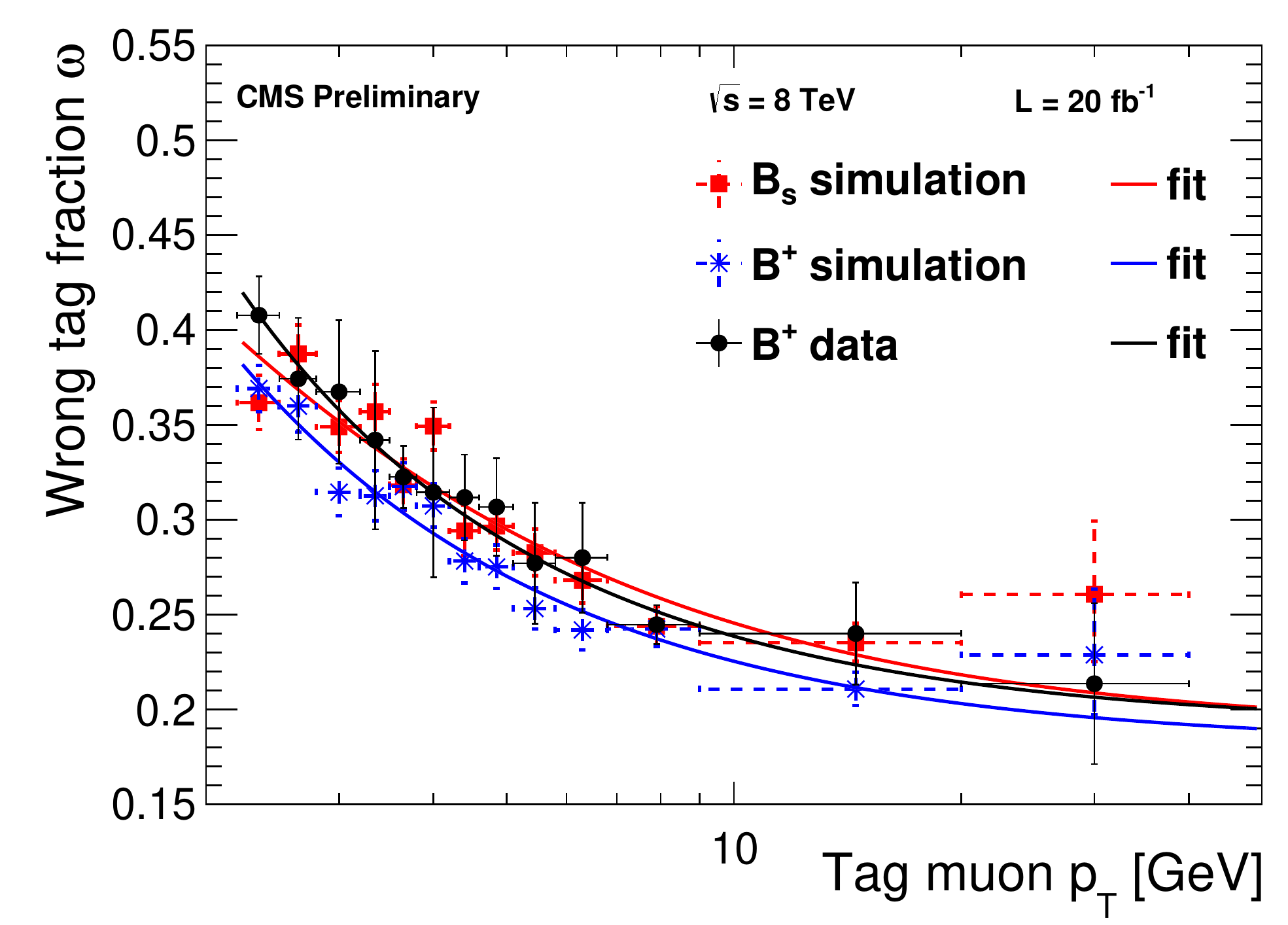}
\includegraphics[width=0.45\textwidth]{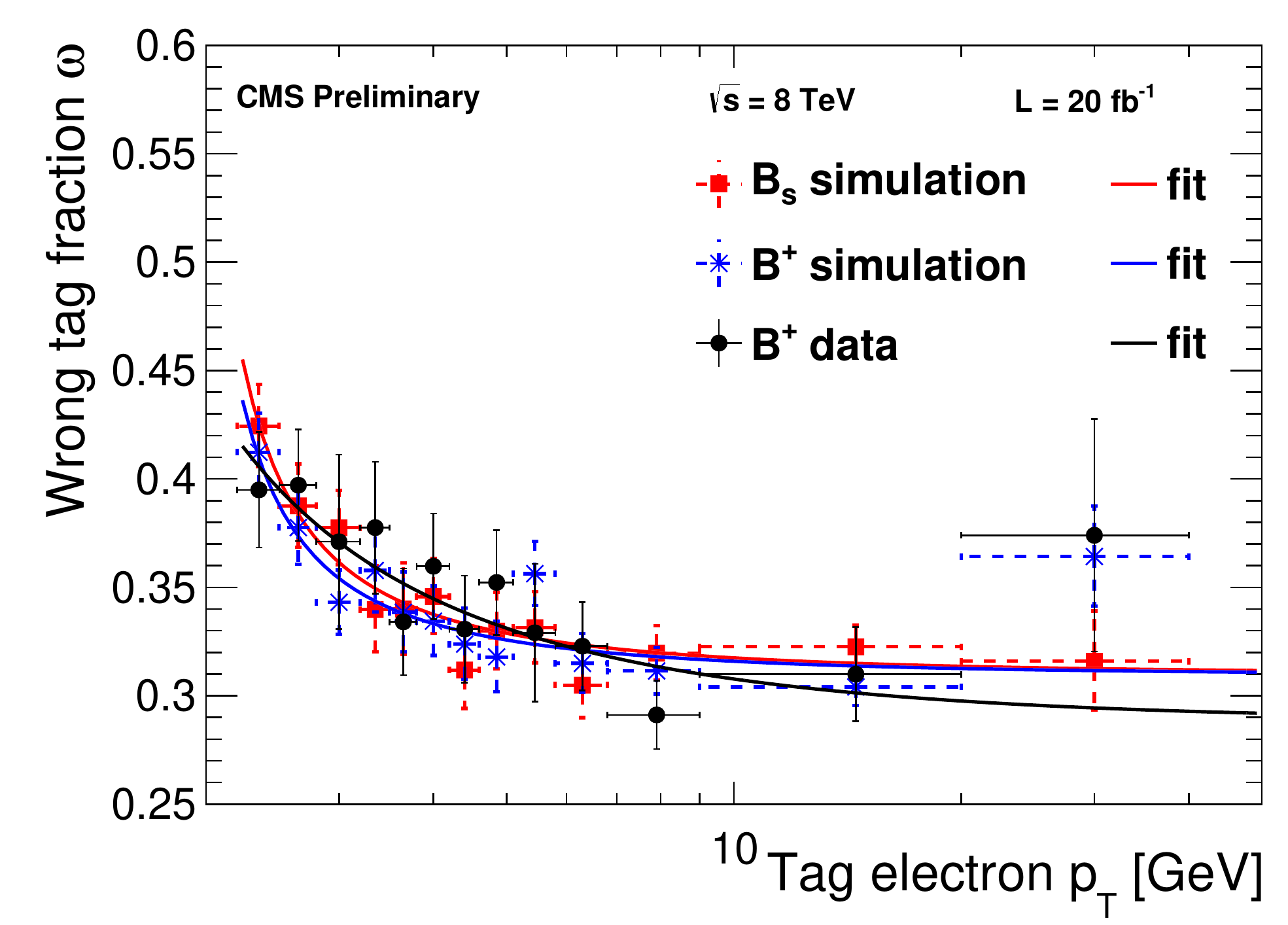} 
\caption{The mistag fraction \wtag as a function of the lepton transverse momentum for muons (top) and electrons (bottom). The points obtained from \Bs simulation (red), \Bp simulation (blue) and \Bp data (black) are shown. The continous lines describe are the relative parametrisations.
 \label{fig:WtagFit2012}}
\end{center}
\end{figure}

A flavour tagging algorithm is used to identify of the flavour of the \Bs meson at production time, improving the sensitivity on \PhiS phase. 
The flavour of the \Bs is inferred on a statistical basis using the properties of the decay products of the opposite side \B hadron, assuming the $\text{b}\bar{\text{b}}$ production process occurred.
%
%
The tagging tool provides the inferred flavour of the \Bs meson and the value of the mistag fraction \wtag, which represents the fraction of incorrectly tagged events.
The tagging efficiency \eff and the mistag fraction \wtag are related to the effective tagging efficiency or tagging power, $\ptag = \eff \left(1 - 2\wtag\right)^2$.
In the present analysis an opposite-side lepton (\muon,\ele) tagger is used. 
For each event the lepton with the highest \pT in the event is selected. 
If no lepton information has been retrieved the tag information is set to zero.
The tag lepton selection is optimized, using \Bs simulations, so that the power of tagging \ptag is maximized separately for electrons and muons. 
%
%
The flavour tagging is measured from data using the self-tagging channel \BpJpsiK. 
The \Bp signal is selected with cuts as similar as possible to those applied to the signal sample.
The tagging performance of simulated \Bp and \Bs events is compared with the \Bp data and found to be consistent. 
In order to increase the sensitivity on \PhiS, the mistag fraction \wtag is binned and parametrized as a function of the transverse momentum of
the lepton, as shown in Fig.~\ref{fig:WtagFit2012}. 
If multiple tag leptons are found in the event, the tagger with the lowest mistag is selected. 
The mistag fraction is assigned to all the tagged events using the parametrisations obtained for electrons and muons in \Bp data. 
The combined tagging performances evaluated on data are $\wtag = (32.2 \pm 0.3)\%$, $\eff = (7.67 \pm 0.04)\%$ and $\ptag = (0.97 \pm 0.03)\%$, where the reported uncertainties are statistical only.

\section{Maximum likelihood fit}\label{sec:fit}
An unbinned maximum likelihood fit to the data is performed by including information on the invariant mass ($m$), proper decay length ($ct$), the three decay angles ($\Theta$) of the reconstructed \Bs candidates, and the proper decay length uncertainty ($\sigma_{ct}$) obtained propagating the uncertainties of the proper decay length measurement. 
From this multi-dimensional fit, the parameters of interest \DeltaGammaS, \PhiS, the \Bs mean lifetime $c\tau$, $|A_{\perp}|^2$, $|A_{0}|^2$, $|A_S|^2$, and the strong phases $\delta_{\parallel}$, $\delta_{\perp}$ and $\delta_{S\perp}$ are determined.
The event likelihood function ${\cal{L}}$ can be represented as described in Eq.~\ref{eqt:FullPDF}, where $L_\text{sig}$ is the PDF that describes the \BsJpsiPhi signal model and $L_\text{bkg}$ describes the background contributions.
\begin{align}
{\cal{L}} &= L_{\text{sig}} +  L_{\text{bkg}} \\ 
L_{\text{sig}} &=  N_S\cdot \left(\tilde{f}(\Theta,\alpha,ct) \otimes G(ct,\sigma_{ct})\cdot \epsilon(\Theta) \right) \cdot \nonumber \\
               & \hspace*{5mm}  P_S(m_{\Bs})\cdot P_S(\sigma_{ct})\cdot P_S(\xi)  \nonumber \\
L_{\text{bkg}} &=  N_{BG}\cdot P_{BG}(\cos\theta_T,\varphi_T)\cdot P_{BG}(\cos\psi_T) \cdot  \nonumber \\
               & \hspace*{5mm} P_{BG}(ct) \cdot P_{BG}(m_{\Bs}) \cdot P_{BG}(\sigma_{ct})\cdot P_{BG}(\xi) \nonumber
\label{eqt:FullPDF}
\end{align}
The PDF $\tilde{f}(\Theta,\alpha,ct)$ is the differential decay rate function defined in Eq.~\ref{eqnarray:decayrate} modified to include the flavour tagging information and the dilution term $(1-2\omega)$.
In the model $\tilde{f}$ the longitudinal phase $\delta_0$ is set to zero, and the difference of phases $\delta_{S}-\delta_{\perp}$ is fitted with a unique variable $\delta_{S\perp}$ to reduce the correlation among the fitted parameters.
Here $\epsilon(\Theta)$ is the angular efficiency function and $G$ is a Gaussian resolution function, which makes use of the event-by-event proper decay length uncertainty $\sigma_{ct}$ scaled by a factor $\kappa$, which is a function of $ct$. 
The $\kappa$ factor is a scale factor introduced to correct the proper decay length uncertainty in order to resemble the actual resolution. 
It is measured in simulated samples assuming that the $\kappa$ factor is the same as in data.
For this assumption a systematic uncertainty is evaluated. 
All the parameters of the PDFs are left free to float in the final fit, unless explicitly stated otherwise. 
The signal mass PDF $P_S(m_{\Bs})$ is given by the sum of three Gaussian functions with a common mean; the two smaller widths, the mean and the fraction of the Gaussians are fixed to the values obtained in a one-dimensional mass fit.
The background mass distribution $P_{BG}(m_{\Bs})$ is described by an exponential function. 
The background proper decay length component $P_{BG}(ct)$ is described by the sum of two exponential functions. 
The angular part of the background PDFs, $P_{BG}(\cos\theta_T,\varphi_T)$ and $P_{BG}(\cos\psi_T)$, are described analytically by a series of Legendre polynomials for $\cos\theta_T$ and $\cos\psi_T$ and sinusoidal functions used for the angle $\varphi_T$. 
For the $\cos\theta_T$ and $\varphi_T$ variables a two-dimensional PDF is used to take into account the correlation among the variables.
The proper decay length uncertainty signal PDF $P_S(\sigma_{ct})$ is a sum of two Gamma functions, where all the parameters are fixed to the values obtained fitting a sample of background-subtracted events. 
The proper decay length uncertainty background PDF $P_{BG}(\sigma_{ct})$ is represented by a single Gamma function, where all the parameters are fixed to the values obtained fitting the mass peak sideband events. 
The $P_S(\xi)$ and $P_{BG}(\xi)$ are the flavour tag decision $\xi$ PDFs which have been obtained from the data sample.

\begin{figure}[htb]
  \centering
  \includegraphics[width=0.404352\textwidth]{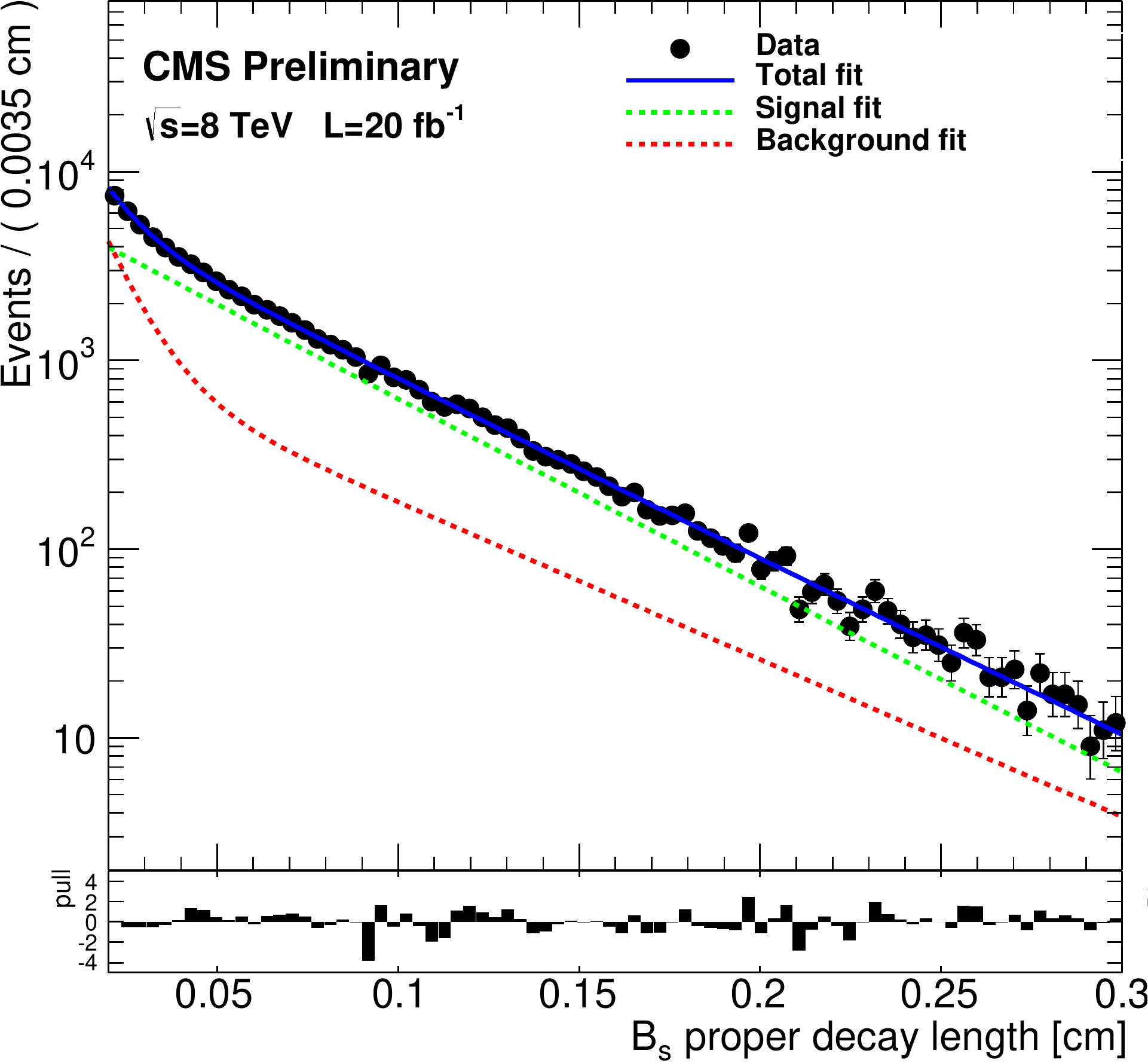}    \\
  \vspace*{5mm}
  \centering
  \includegraphics[width=0.433836\textwidth]{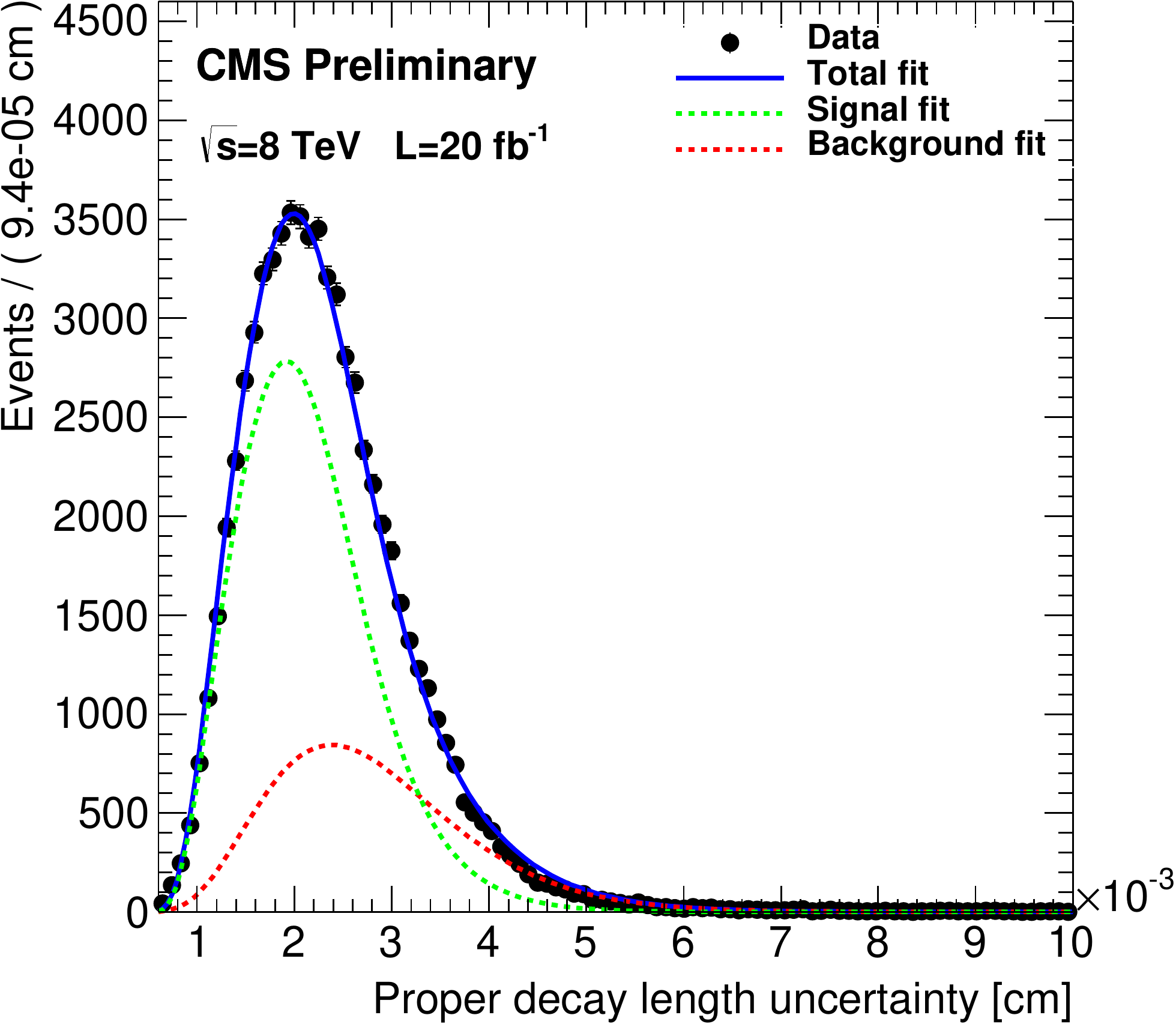}    
  \centering
  \caption{The proper decay length distribution (top) and the proper decay length uncertainty (bottom) of the \Bs candidates. The full line is a fit to the data (solid markers), the dashed green line is the fitted signal and the dashed red line is the fitted background. For the proper decay length distribution the pull between the histogram and the fitted function is displayed in the histogram below. }
  \label{fig:lifetime}
\end{figure}

\begin{figure}[htb]
  \centering
  \includegraphics[width=0.235\textwidth]{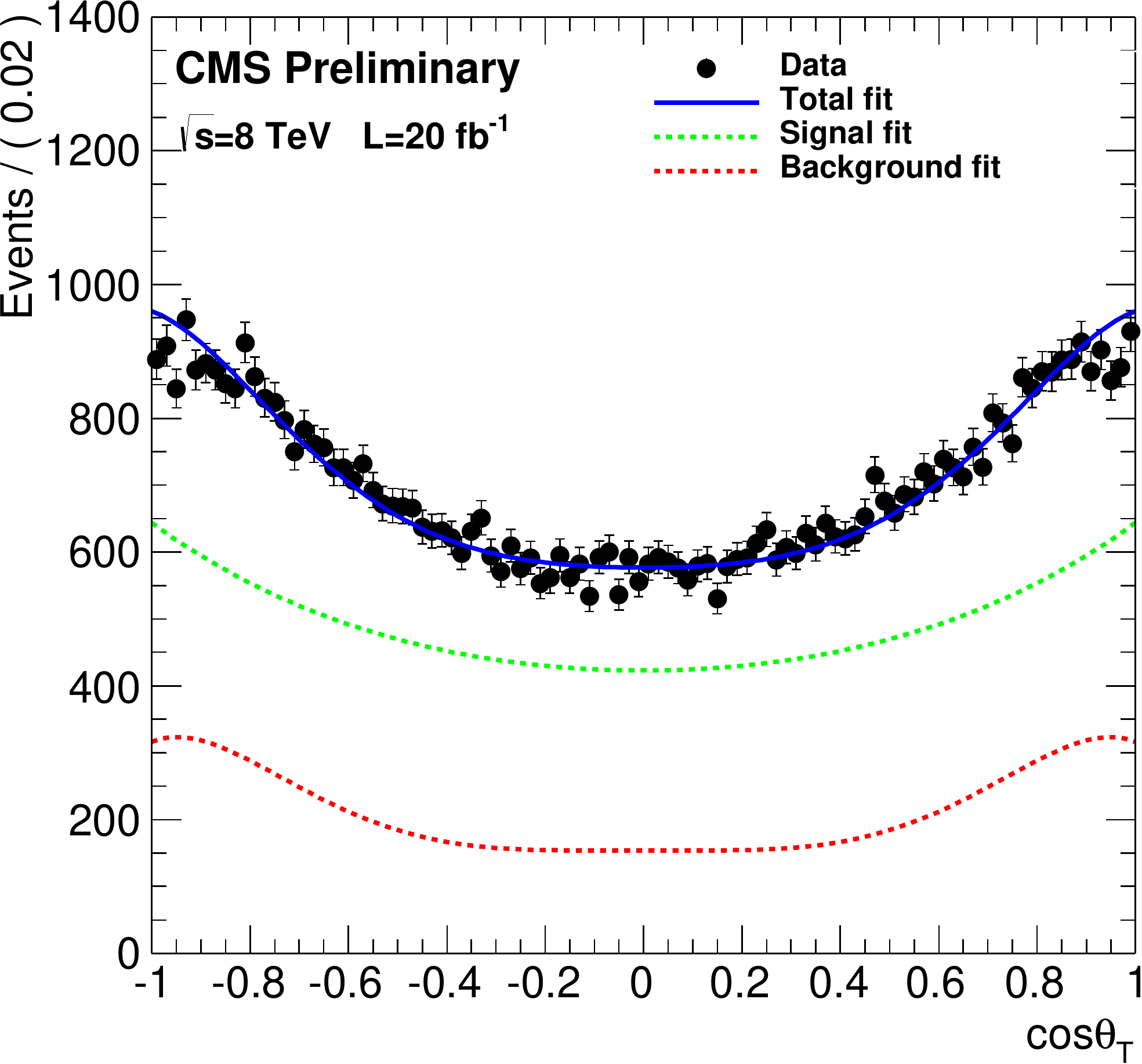} \\
  \centering
  \includegraphics[width=0.235\textwidth]{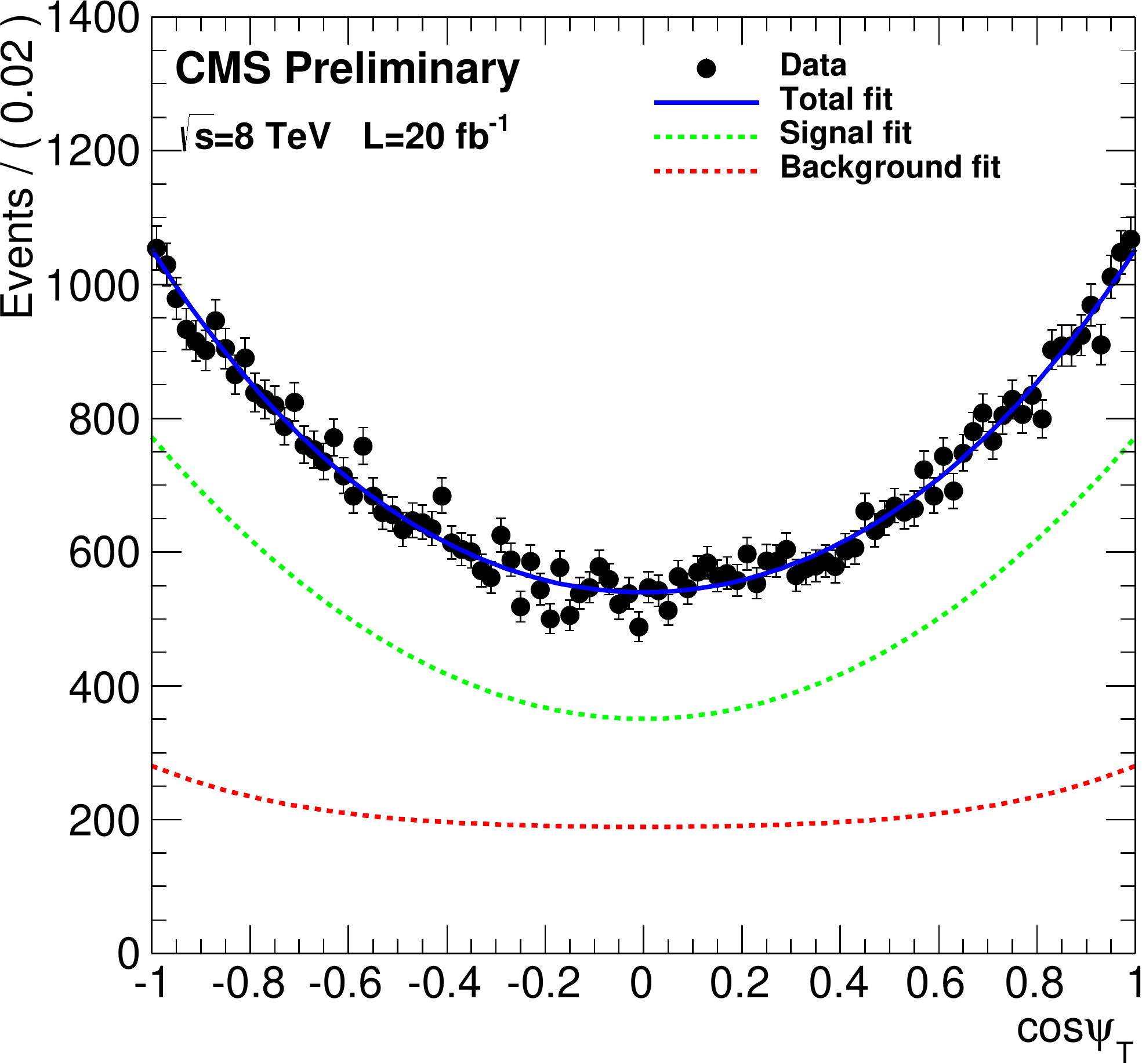}   
  \includegraphics[width=0.235\textwidth]{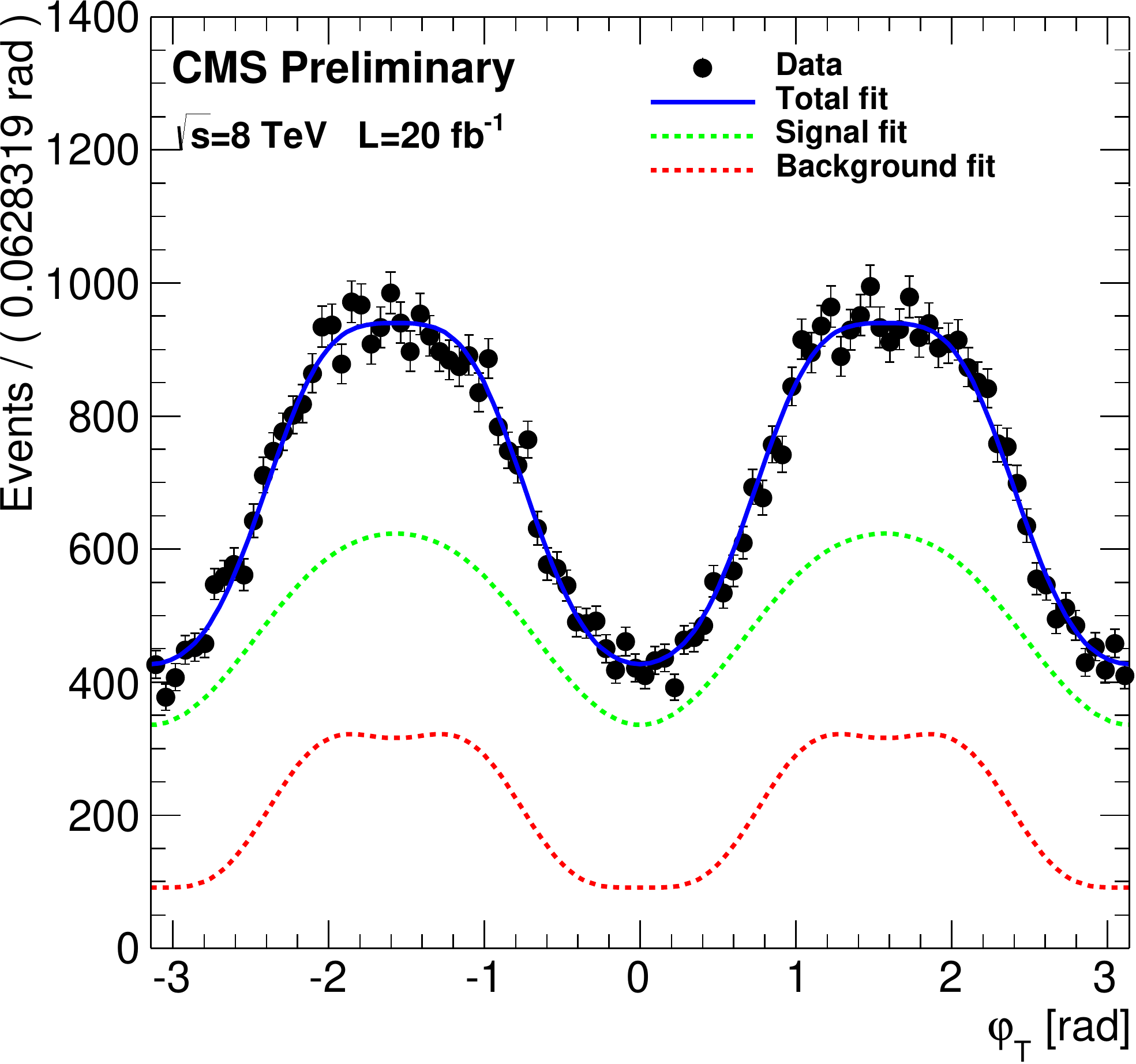}    
  \centering
  \caption{The angular distributions ($\cos\theta_T$, $\cos\psi_T$, $\varphi_T$) of the \Bs candidates. The full line is a fit to the data (solid markers), the dashed green line is the fitted signal, and the dashed red line is the fitted background.}
  \label{fig:angles}
\end{figure}
 
\section{Results}\label{sec:res}
\begin{figure}[hbt]
\begin{center}
\includegraphics[width=0.45\textwidth]{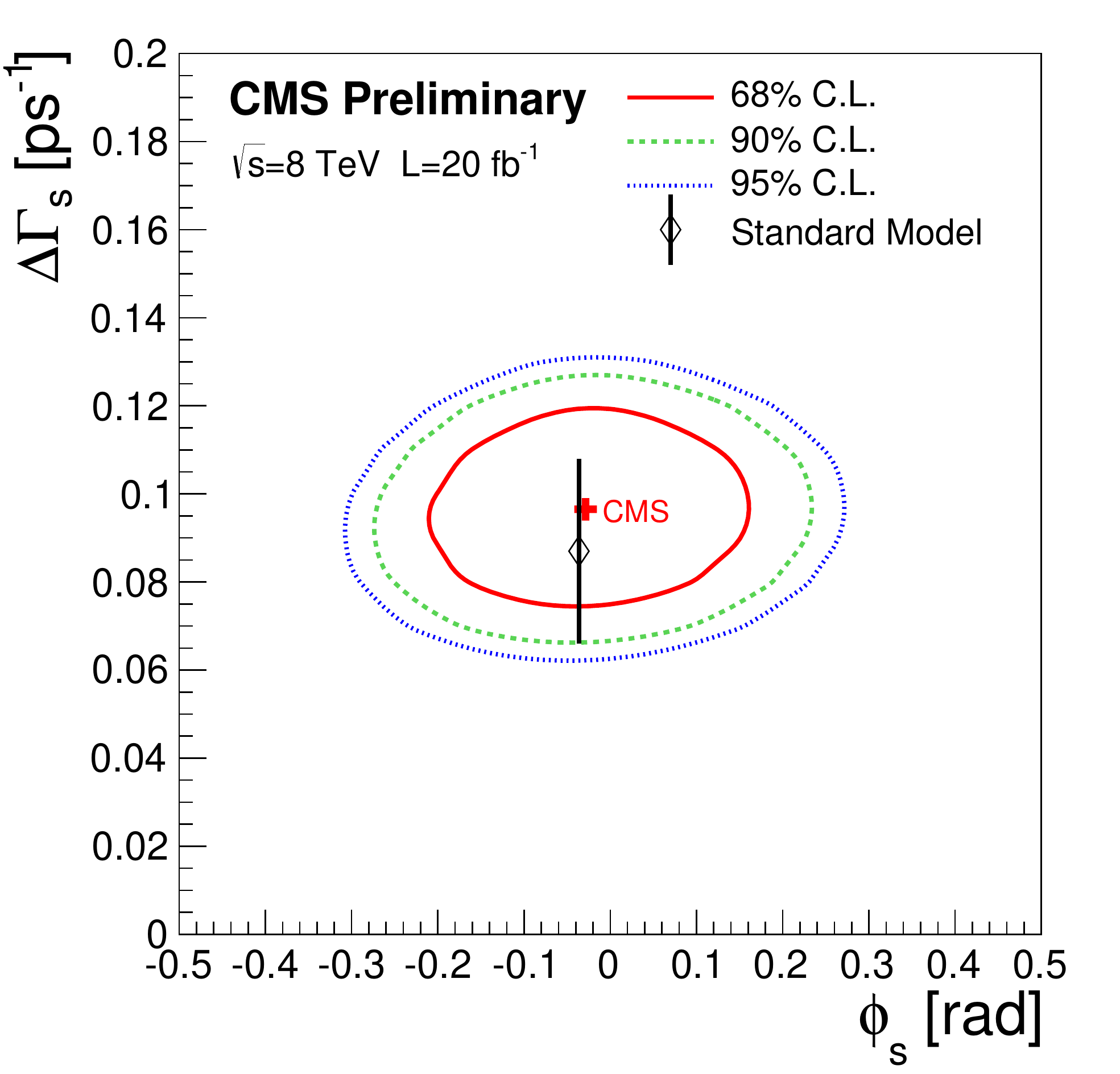} 
\end{center}
\caption{The 68\%, 90\% and 95\% C.L. contours in the \DeltaGammaS versus \PhiS plane, together with the SM fit prediction. Uncertainties are statistical only.}
\label{fig:2dcontour}
\end{figure}

The fit is applied to the sample of $70\,000$ events ($49\,000$ signal candidates and $21\,000$ background events), selected in the mass range $[5.24-5.49]~\GeV$ and proper decay length range $[200-3\,000]~\micron$. 
The $\Delta m_s$ has been constrained in the fit to the current world average value $(17.69\pm 0.08)\times 10^{12}~\hbar/\mathrm{s}$~\cite{bib:Beringer} by taking a Gaussian distribution centred on the world average with the uncertainty as the width. 
No direct CP violation is assumed for this measurement, and therefore $|\lambda|$ is set to one, consistent with the results in Ref.~\cite{bib:Aaij_14}.
The \DeltaGammaS is constrained to be positive as in Ref.~\cite{bib:Aaij_12}.
The observable distributions and the fit projections are shown in Figs.~\ref{fig:mass}, \ref{fig:lifetime}, and \ref{fig:angles}. 
The 68\%, 90\% and 95\% Confidence Level (C.L.) likelihood contours of the fit for \PhiS and \DeltaGammaS are shown in Fig.~\ref{fig:2dcontour}.
The fit results are presented in Table~\ref{table:datatagged}, where the uncertainties are statistical only.
 
\begin {table}[hb]
\centering
  \begin{tabular}{ l  r@{$\,\pm\,$}l }
    \toprule
    Parameter                                   & \multicolumn{2}{c}{Fit result}      \\
    \midrule
    $|A_0|^2$                                   & 0.511 & 0.006  \\
    $|A_S|^2$                                   & 0.015 & 0.016  \\
    $|A_{\perp}|^2$                             & 0.242 & 0.008  \\
    $\delta_{\parallel}~\mathrm{[rad]}$         & 3.48 & 0.09  \\
    $\delta_{S\perp}~\mathrm{[rad]}$            & 0.34 & 0.24  \\
    $\delta_{\perp}~\mathrm{[rad]}$             & 2.73 & 0.36  \\
    $c\tau~[\mu\mathrm{m}]$                     & 447.3 & 3.0  \\
    $\DeltaGammaS~[\ps]$                        & 0.096 & 0.014  \\
    $\PhiS~[\rad]$                              & -0.03 & 0.11  \\
    \bottomrule
  \end{tabular}
\caption{Results of the fit to the 2012 data. Only the statistical uncertainties are shown.}
\label{table:datatagged}
\end{table}
 
\section{Systematics}\label{sec:syst}
The systematic uncertainties are summarized in Table~\ref{table:systematics}. 
The uncertainties of the \PhiS and \DeltaGammaS results are dominated by statistical uncertainties.
The systematic uncertainty associated with the hypothesis of a flat proper decay length efficiency is evaluated by fitting the data with the a proper decay length efficiency which takes into account a small contribution of the decay length significance cut at small $ct$ and a first order polynomial variations at high $ct$.
The uncertainties associated with the variables of threedimensional angular efficiency function $\cos\theta_T$, $\cos\psi_T$, and $\varphi_T$ are propagated to the fit results by varying the corresponding parameters within their statistical uncertainties and accounting for the covariances among the parameters. 
The systematic uncertainty due to a small discrepancy in the kaon \pT spectrum between the data and the simulations is evaluated by reweighting the simulated kaon \pt spectrum to agree with the data.
The intrinsic biases of the fit model are taken into account as a systematic effect.
The uncertainty in the proper decay length resolution associated with the proper decay length uncertainty scale factor $\kappa$ is propagated to the results. 
Since the $\kappa(ct)$ factors are obtained from simulation, the associated systematic uncertainty is assessed by using a sample of prompt \Jpsi decays obtained with an unbiased trigger and comparing them to similarly processed simulated data.
The likelihood does not contain a PDF model for the mistag distribution, therefore the systematic uncertainty arising from this source is estimated. 
The systematic uncertainty due to tagging is assessed by propagating the statistical and systematic uncertainty of the \wtag parametrisation to the results.
The various hypotheses that have been assumed when building the likelihood function are tested by generating simulated pseudo-experiments with different hypotheses in the generated samples and fitting the samples with the nominal likelihood function.
Finally the $|\lambda| = 1$ hypothesis is tested by leaving that parameter free in the fit. 
The obtained value of $|\lambda|$ agrees with one within one standard deviation. 
The differences found in the fit results with respect to the nominal fit are used as systematic uncertainties.
\begin {table*}[htb]
\footnotesize
\centering
  \begin{tabular}{@{\hskip 0pt} l@{\hskip 10pt} c@{\hskip 5pt}  c@{\hskip 5pt}  c@{\hskip 5pt}  c@{\hskip 5pt}  c@{\hskip 5pt}  c@{\hskip 5pt}  c@{\hskip 5pt}  c@{\hskip 5pt}  c @{\hskip 0pt}}
    \toprule
    Source of uncertainty             & $|A_0|^2$  &  $|A_S|^2$  & $|A_{\perp}|^2$  & $\Delta\Gamma_\mathrm{s}~[\mathrm{ps}^{-1}]$ &  $\delta_{\parallel}~[\mathrm{rad}]$ &   $\delta_{S\perp}~[\mathrm{rad}]$ &   $\delta_{\perp}~[\mathrm{rad}]$  & $\phi_\mathrm{s}~[\mathrm{rad}]$ & $c\tau~[\mu \mathrm{m}]$ \\
    \midrule
    Statistical uncertainty             & 0.0058        & 0.016           & 0.0077                  & 0.0138                                         & 0.092                                    & 0.24                                      & 0.36                                        & 0.109                              & 3.0\\           
    \midrule
    Angular efficiency                   & 0.0060        &  0.008          &  0.0104                 & 0.0021                                         & 0.674                                    & 0.14                                      & 0.66                                        & 0.016                 & 0.8 \\
    $|\lambda|$ as a free parameter     & 0.0001        &  0.005         &  0.0001                 & 0.0003                                         & 0.002                                     & 0.01                                     & 0.03                                         & 0.015               & - \\
    Model bias                          & 0.0008       & -                    & -                             &  0.0012                                         & 0.025                                    & 0.03                                      & -                                              &  0.015                & 0.4 \\
Kaon $p_\mathrm{T}$ re-weighting& 0.0094   &  0.020         &  0.0041                 & 0.0015                                         & 0.085                                     & 0.11                                     & 0.02                                        & 0.014               & 1.1 \\
    Proper decay length resolution         & 0.0009       & -                   & 0.0008                  & 0.0021                                         & 0.004                                    & -                                             & 0.02                                         & 0.006                & 2.9 \\
    PDF modelling assumptions  &  0.0016      &  0.002           &  0.0021                &  0.0021                                     &  0.010                                    &  0.03                                   &  0.04                                       & 0.006                &     0.2\\
    Flavour tagging                        & -                  &  -                   & -                              & -                                               & -                                            & -                                             & 0.02                                        & 0.005                & - \\
    Background mistag modelling    & 0.0021 & -                    & 0.0013                  & 0.0018                                         & 0.074                                   & 1.10                                       & 0.02                                        & 0.002              & 0.7\\
    Proper decay length efficiency            & 0.0015        &  -                   &  0.0023                 & 0.0057                                         & -                                             & -                                             & -                                              & 0.002               & 1.0 \\
    \midrule
    Total systematics                     &   0.0116      &  0.022         &  0.0117                  & 0.0073                                       &  0.684                                  & 1.12                                      & 0.66                                        &  0.032                 & 3.5\\ \bottomrule
  \end{tabular}
\caption{Summary of the uncertainties. If no value is reported, then the systematic uncertainty is negligible with respect to the statistical and other systematic uncertainties. The total systematic uncertainty is the square root of sum of squares of the listed systematic uncertainties.}
\label{table:systematics}
\end{table*}
\section{Conclusions}\label{sec:concl}
Using the 2012 CMS data approximately 49000 \Bs signal candidates were reconstructed and used to accurately measure the weak phase \PhiS and the decay width difference \DeltaGammaS. 
The analysis was performed by using opposite-side lepton tagging of the \Bs flavour at the production time. 
Both muon and the electron tags were used.
The measured values for the weak phase and the decay width difference between the \Bs mass eigenstates are:

\begin{align}
\phi_\mathrm{s} &= -0.03 \pm 0.11~\mathrm{(stat.) \pm 0.03~(syst.)~rad} \\ 
\Delta\Gamma_\mathrm{s}&=  0.096 \pm 0.014 ~\mathrm{(stat.)} \pm 0.007~\mathrm{(syst.)~ps}^{-1} 
\end{align}

where the uncertainties of the \PhiS and \DeltaGammaS results are dominated by statistical uncertainties.
The value of \PhiS is in agreement with the previous measurements and with the SM fit prediction, and \DeltaGammaS is confirmed to be non-zero. 
%
%

\end{document}